# *Penrose Model potential, compared with Coleman-Weinberg Potential for early universe scalar evolution*

A.W. Beckwith

*abeckwith@UH.edu*

## Abstract

We present evidence in terms of a D'Alembertain operator acting on a scalar field minus the first derivative of a potential system, with respect to an inflaton scalar field, that the Penrose model as outlined as an alternative to cosmological big crunch models gives us emergent behavior for an inflaton scalar field in early universe cosmological models. This is in contrast to the Coleman-Weinberg potential which in low temperature conditions is always presenting almost non existent emergent scalar fields. This permits us to state that Penrose's cyclic universe model in its initial conditions gives us scalar field dynamics consistent with emergent scalar fields which die out quickly as temperature drops after the onset of inflation. We make no attempt to find the particulars of the conformal mapping which allows the alternative to the big crunch Penrose (2007) lectured upon in the inaugural meeting of the IGC at Penn State..

**Key words** : **Inflaton scalar field, Penrose cyclic model, Coleman-Weinberg potential**

**PACS**: 98.80.-k

# Introduction

We begin first with describing the relevant equation of motion involving the D'Alembertain operation upon a scalar field minus the derivative of a potential system for the scalar potentials with respect to the inflaton. With regards to the Penrose potential, we find largely due to the existence of a scalar mass term added which disappears due to high temperature conditions a procedural addition to permit the inflaton field in high temperatures to nucleate and then to subsequently fade out with dropping temperature conditions as occurs in the aftermath of inflation. The subsequent alternative to the big crunch which Penrose (2007) lectured upon in the IGC inaugural meeting involves a conformal mapping which the author has not had time to work on, yet. We can say this though; for a nucleation of an emergent scalar potential, we would first need a low temperature regime which heated up. The author after establishing the validity of Penrose's (2007) procedure for emergent scalar space time in initial phases refers to a scheme involving worm hole transfer of thermal input into the initial conditions of inflation. The author then closes with some observations as to necessary and sufficient conditions for the formation of the conformal mapping collation of gravitational radiation energy which Penrose (2007) alluded to in his IGC talk, involving the ripples in the pond effect, where gravitational radiation from Black holes, collected in the aftermath of Hawking radiation is collated into a new big bang. Penrose (2007) in conversations with the author at Penn state alluded to the Hawking radiation effect as a causal effect which permits a necessary condition for the formation of necessary conditions for the conformal mapping of many black hole Hawking radiation releases into a singular big bang. The author, without trying to ascertain the particulars of this conformal mapping discusses what he thinks constitutes necessary conditions for the formation of such a transformation.

# Preliminaries: The D'Albembertain operation in an equation of motion for emergent scalar fields

We begin with the D'Albertain operator as part of an equation of motion for an emergent scalar field. We then refer to the Penrose potential which the author gleaned from the presentation given by Penrose (2007) at the inauguration of the IGC center, at Penn States, August 2007. This potential is then shown, with an initial assumption of Euclidian flat space for computational simplicity to account for, in a high temperature regime an emergent non zero value for the scalar field $\phi$ due to a zero effective mass, at high temperatures. When the mass approaches far lower values, it, a non zero scalar field re appears. Leading to $\phi \xrightarrow[T \to 2.7^0 Kelvin]{} \varepsilon^+ \approx 0^+$ as a vanishingly small contribution to cosmological evolution, as we have non zero effective mass for low temperatures. This dynamic will be, in the next section, compared to what happens in the Coleman-Weinberg effective potential which is almost always mandating $\phi \xrightarrow[T \to 2.7^0 Kelvin]{} \varepsilon^+ \approx 0^+$. The author is aware that the Euclidian flat space hypothesis, at the initial phases of nucleation, and indeed a flat space metric is likely not accurate at the initial phases of nucleation. Needless to say, this simple model identifies the initial emergent scalar conditions for what the author views as a quintessence contribution to cosmological evolution. This section also speculates as to what is needed for thermal input to create circumstances for emergent scalar potentials. One model is the discussion by Penrose (2007) as to a non standard solution of innumerable black holes contributing to a future big bang. This is done in terms of a perpetually expanding universe whose matter, as it cools is collated into innumerable black holes. As the thermal temperature drops below the 2.7 K limit, Penrose (2007) suggests that a conformal transformation makes a bridge between these innumerable black holes releasing material via a Hawking radiation procedure to be collated into a future big bang. The author frankly has conceptual difficulties with this picture, but will bring up what he views as necessary conditions for such a mapping to take place. The other picture, brought up in the next section which the author favors is one of a wormhole transferal of thermal radiation from a prior contracting universe into this present universe, via a pseudo time dependent Wheeler De Witt equation, as given by Crowell (2005). The solution so obtained has symmetries which will be discussed in passing, but in certain respects has a partial WKB character which will be remarked upon by the author.

Let us now begin to initiate how to model the Penrose quintessence scalar field evolution equation. To begin, look at the flat space version of the evolution equation, which Penrose proposed at the IGC inaugural conference. It reads as

$$\ddot{\phi} - \nabla^2 \phi + \frac{\partial V}{\partial \phi} = 0 \tag{1}$$

This is, in the Friedman – Walker metric using the following as a potential system to work with, namely:

$$V(\phi) \sim -\left[\frac{1}{2} \cdot \left(M(T) + \frac{\Re}{6}\right)\phi^2 + \frac{\tilde{a}}{4}\phi^4\right] \equiv -\left[\frac{1}{2} \cdot \left(M(T) + \frac{\kappa}{6a^2(t)}\right)\phi^2 + \frac{\tilde{a}}{4}\phi^4\right] \tag{2}$$

This is pre supposing $\kappa \equiv \pm 1, 0$, that one is picking a curvature signature which is compatible with an open universe. That means $\kappa = -1, 0$ as possibilities. Picking the closed universe is not compatible with the assumptions of Penrose's concept which is for an open universe, where Black holes collate matter far into the future and have the material so collected released into a new big bang. While not necessarily endorsing the last part of Penrose's supposition, we will look at the $\kappa = -1, 0$ values so as to determine what a good emergent scalar field match is. When we do so we find that we need to have an anzatz for the scalar field to work with. We begin with.

$$\ddot{\phi} - \nabla^2 \phi + \frac{\partial V}{\partial \phi} = 0 \Rightarrow \phi^2 = \frac{1}{\tilde{a}} \cdot \left\{c_1^2 - \left[\alpha^2 + \frac{\kappa}{6a^2(t)} + M(T)\right]\right\} \Leftrightarrow \phi \equiv e^{-\alpha \cdot r} \exp(c_1 t) \tag{3}$$

We find the following as far as basic phenomenology, namely

$$\phi^2 = \frac{1}{\tilde{a}} \cdot \left\{c_1^2 - \left[\alpha^2 + \frac{\kappa}{6a^2(t)} + M(T)\right]\right\} \xrightarrow[M(T \sim high) \to 0]{} \phi^2 \neq 0 \tag{4}$$

$$\phi^2 = \frac{1}{\tilde{a}} \cdot \left\{c_1^2 - \left[\alpha^2 + \frac{\kappa}{6a^2(t)} + M(T)\right]\right\} \xrightarrow[M(T \sim Low) \neq 0]{} \phi^2 \approx 0 \tag{5}$$

Our next task will be to come up with a suggestion as to how there could be a switch from Eqn. (4) to Eqn. (5). One suggestion as brought up independently of the Penrose model itself has to do with work the author performed in a different model which if super imposed upon how Eqn. (4) and Eqn. (5) inter relate give us the very real possibility if we initially have a pre inflationary state of low temperature that the worm hole model as mentioned in the next section could give us emergent quintessence fields which damp out quickly. After we establish this in the next section, will be a comparison with the Coleman Weinberg potential which almost always gives quintessence behavior as given by Eqn.(5).

## Emergent space time. As generated via a wormhole

Lorentzian worm holes have been modeled quite thoroughly, and Visser (1995) states that the wormhole solution does not have an event horizon hiding a singularity, i.e. there is no singularity in the wormhole held open by dark energy. So being the case, the only case a wormhole could form would be as a bridge between a prior to a present universe, which is what Crowell (2005) refers to in his reference on quantum fluctuations of space time tome which uses a pseudo time like space co ordinate to a modified Wheeler – De Witt equation for a bridge between two universes. We add in another caveat, that the worm hole solution is dominated by a huge vacuum energy value. This leads to the following situation, which we present here:

To model this, we use results from Crowell (2005) on quantum fluctuations in space time which gives a model from a pseudo time component version of the Wheeler De Witt equation, with a use of the Reinssner-Nordstrom metric to help us obtain a solution which passes through a thin shell separating two space times. The radius of the shell, $r_0(t)$ separating the two space times is of length $l_P$ in approximate magnitude, leading to a domination of the time component for the Reissner – Nordstrom metric

$$dS^2 = -F(r) \cdot dt^2 + \frac{dr^2}{F(r)} + d\Omega^2 \tag{6}$$

This has:

$$F(r) = 1 - \frac{2M}{r} + \frac{Q^2}{r^2} - \frac{\Lambda}{3} \cdot r^2 \xrightarrow[T \to 10^{32} Kelvin \sim \infty]{} -\frac{\Lambda}{3} \cdot (r = l_P)^2 \tag{7}$$

This assume that the cosmological vacuum energy parameter has a temperature dependence as outlined by Park (2003) leading to if

$$\frac{\partial F}{\partial r} \sim -2 \cdot \frac{\Lambda}{3} \cdot (r \approx l_P) \equiv \eta(T) \cdot (r \approx l_P) \tag{8}$$

As a wave functional solution to a Wheeler De Witt equation bridging two space times. This solution bridging two space times is similar to that being made between these two space times with 'instantaneous' transfer of thermal heat ,as given by Crowell (2005)

$$\Psi(T) \propto -A \cdot \{\eta^2 \cdot C_1\} + A \cdot \eta \cdot \omega^2 \cdot C_2 \tag{9}$$

This has $C_1 = C_1(\omega, t, r)$ as a pseudo cyclic and evolving function in terms of frequency, time, and spatial function, with the same thing describable about $C_2 = C_2(\omega, t, r)$ with $C_1 = C_1(\omega, t, r) \neq C_2(\omega, t, r)$. The upshot of this is that a thermal bridge between a shrinking prior universe, collapsing to a singularity, and an expanding universe expanding from a singularity exits, with an almost instantaneous transfer of heat with terms dominated by $\eta(T)$ exits, and is forming a necessary and sufficient condition for the thermal heat flux. We get that from this is in part due to the identification which we will explicitly state, namely that by assuming that the absolute value of the five dimensional 'vacuum state' parameter varies with temperature T, as Beckwith (2007) writes

$$|\Lambda_{5-\dim}| \approx c_1 \cdot (1/T^\alpha) \tag{10}$$

in contrast with the more traditional four dimensional version of the same, minus the minus sign of the brane world theory version The five dimensional version is actually connected with Brane theory, and higher dimensions, whereas the four dimensional is linked to more traditional De Sitter space time geometry, as given by Park(2003)

$$\Lambda_{4-\text{dim}} \approx c_2 \cdot T^\beta \tag{11}$$

This is such that If one looks at the range of allowed upper bounds of the cosmological constant, we have that the difference between what Barvinsky (2006) recently predicted, and Park (2003) specifying an upper limit as of 2003, based upon thermal input is a give away that a phase transition is occurring at or before Planck's time . This allows for a brief interlude of quintessence We should note that this is assuming that a release in gravitons occurs which leads to a removal of graviton energy stored contributions to this cosmological parameter, with $m_P$ as the Planck mass, i.e. the mass of a black hole of 'radius' on the order of magnitude of Planck length $l_P \sim 10^{-35}$ m. This leads to Planck's mass $m_P \approx 2.17645 \times 10^{-8}$ kilograms, as alluded to by Barvinsky (2006)

$$\Lambda_{4-\text{dim}} \propto c_2 \cdot T \xrightarrow{graviton-production} 360 \cdot m_P^2 << c_2 \cdot \left[T \approx 10^{32} K\right] \tag{12}$$

Needless to say, right after the gravitons are released one still is seeing a drop off of temperature contributions to the cosmological constant .Then we can write, for small time values $t \approx \delta^1 \cdot t_P, 0 < \delta^1 \leq 1$ and for temperatures sharply lower than $T \approx 10^{12} Kelvin$, Beckwith (2007), where for a positive integer $n$

$$\frac{\Lambda_{4-\text{dim}}}{|\Lambda_{5-\text{dim}}|} - 1 \approx \frac{1}{n} \tag{13}$$

The transition outlined in Eqn. (12) above has a starting point with extremely high temperatures given by a vacuum energy transferal between a prior universe and our present universe, as outlined by Eqn. (7) and Eqn. (8) above, whereas the regime where we look at an upper bound to vacuum energy in four dimensions is outlined in Eqn. (13) above, so that eventually we can model the behavior of scalar fields as being transformed from cyclic behavior with an imaginary component to a purely real valued scalar equation as given by the next sections argument. We then conclude with a proof of the short term behavior of this quintessence scalar field, in a manner which makes reference to both Eqn. (12) and Eqn (13) above.
.

## Why quintessence scalar fields damp out

The origins of this methodology lie in looking at first the initial phases of how the Einstein equations evolve, with the assumption made for the sake of simplicity that we can model part of the contributions to the Einstein stress tensor via a Casmir plate treatment made as an approximation to initial 'domain wall' treatment of stored energy in the initial phases of nucleation of a vacuum energy. Let us first review the typical implications as to how we get the Einstein field equations. We will then proceed to consider how the Einstein equation, with a bit of emphasis as to the proposal as to the evolution of the vacuum energy actually gives credence to the necessity of short term existence of the quintessence scalar field. The proposal so outlined heavily depends upon a huge vacuum energy being dominant in the derived Einstein field equations, with the combined stress energy tensor contributions set equal to zero. The evolution of the scale factor would be in tandem with adding a new term to the metric $g_{u,v}$, due to adding in a scale factor contribution to actually read as given by Moffat (2002):

$$\tilde{g}_{uv} \equiv g_{u,v} - \vartheta \cdot \left(\partial_u \phi \ \ \partial_v \phi\right) \tag{14}$$

Let us now look at the Stress tensor in General Relativity. We get, as a take off from Birrell (1984)

$$T_{u,v} = \frac{2}{\sqrt{-g}} \cdot \frac{\delta S}{\delta g_{u,v}} \equiv 0,$$

$$S \equiv S_g + S_M,$$

$$T_{u,v}\Big|_{S \to S_M} = \frac{4\pi AG}{a}\Big|_{a \to \varepsilon^+} \quad (15)$$

$$\Rightarrow R_{u,v} - \frac{g_{uv} \cdot R}{2} + \Lambda_{4-\dim} g_{u,v} = -8\pi G T_{u,v}$$

$$\Rightarrow \Lambda_{4-\dim}\Big|_{initial} \cdot \left[g_{0,0} - \vartheta \cdot [\partial_0 \phi]^2\right] = \frac{4\pi AG}{\varepsilon^+}$$

The last segment on the bottom of Eqn. (15) captures the dynamics of the scalar field interaction. The statement made above in particular relies upon the following dynamic made in a short time interval. Namely, assume that the right hand side of the Casmir plate separation, as written up by De Witt (1979) would be relatively constant as the separation of the domain walls of an initial vacuum state configuration became miniscule in size

$$\Lambda_{4-\dim}\Big|_{initial} \cdot \left[g_{0,0} - \vartheta \cdot [\partial_0 \phi]^2\right] \xrightarrow[t \to t* < Planck\ time]{} \textbf{constant} \quad (16)$$

In such a short time interval, we would have as the initial cosmological vacuum energy went up, a corresponding drop in the $(\partial_0 \phi) \neq 0 \xrightarrow[t < Planck\ time]{} 0^+$. Either the scalar quintessence field would be a constant, or it would go to zero. Considering the non spherical geometry of early brane world geometry, the easiest way to get uniformly consistent criteria would be to have the quintessence scalar field itself rapidly damp out even if we write having the distance of the separation between early vacuum state geometry, designated as $a$ go nearly to zero. But having the left hand side of $(\partial_0 \phi) \neq 0 \xrightarrow[t < Planck\ time]{} 0^+$ limit initially not zero would be, in tandem with Eqn (16) above a strong lead as to non zero, time changing quintessence fields being a measure of quantum entanglement.

## Comparison with the Coleman-Weinberg low temperature potential for scalar fields

We consider a model of a potential, as given by Coleman-Weinberg which is primarily configured for low temperature regimes. This comes out to be as follows, Kurioukidis (2004), as given by:

$$V(\phi) \sim V(0) + B\phi^4 \cdot \left[\ln\left(\frac{\phi^2}{\sigma^2}\right) - 1/2\right] \quad (17)$$

If the scalar potential is in itself tending to be small, we find that we obtain a logarithmic expansion which, if we use the other assumptions used in filling out Eqn. (1) above:

$$\phi^2 \cong \frac{c_1^2 - \alpha^2}{4B \cdot \left[1 + \sum_{j=2}^{N} \frac{1}{j}\right]} \xrightarrow[N \to \infty]{} 0^+ \quad (18)$$

The more terms are taken in looking at $\ln\left(\frac{\phi^2}{\sigma^2}\right) \sim \sum_{j=1}^{\infty} (-1)^{j+1} \cdot \frac{[(\phi/\sigma) - 1]^j}{j}$, the quicker Eqn (18) converges to zero. I.e. we have an almost instantaneous collapse of the scalar field $\phi$ to zero, which is in effect saying that the Coleman-Weinberg potential is good for duplicating the results of Eqn. (5) only.

# What can be said about a conformal mapping of Hawking radiation leakage from black holes being collated into material for a new 'big bang'?

First what do we mean in this situation by the Hawking radiation effect? In this we are referring to what happens to the life time of such entities even if the mass drops below a typical threshold. As an example:

$$\tau \sim \frac{1}{M_*}\left(\frac{M_{BH}}{M_*}\right)^{(n+3)/(n+1)} \tag{19}$$

Here, the value of $M_*$ can be as low as a few TeV, and the formula can give time scales on the order of "new Planck time" $\sim 10^{-26}$ s, while n can be of any dimension needed. The problems come up with supposed Black holes with life times greater than the typical values assumed for the lifetime of the universe. i.e. Penrose stated that in his model that Black holes would be expected to collect matter far into the future, and that Hawking radiation would leak the collected 'material' in a fashion which would collectively be re constituted for a new 'big bang'. Penrose dubbed his collection of Hawking radiation for innumerable black holes a so called "ripple in the pond" effect which would in its own way able to collate material for a new big bang. Ida, Oda, and Park (2003) give a Hawking temperature which is strongly dependent upon dimensionality, but which affirms roughly that black hole production and evaporation in extra dimension scenarios with TeV scale gravity will still keep much of the qualitative features presented in Eqn (19) above. This then leads us to the following question. What conformal mapping exists which collates material from a Hawking temperature? Ida, Oda, and Park in their article's Eqn. (27) have a dimensionally dependent expression for Hawking temperature, and the Hawking temperature, T, is such that it in general as Klaus Kiefer notes (2002) there is a relationship for an interplay between entropy and a Hawking's temperature T along the lines of, if the surface gravity $K_{surface} \sim GM/R_0^2$, where the denominator is roughly the radii of a black hole, and, as given by Kiefer (2002)

$$T_{BH} \cdot dS_{BH} \sim \lfloor K_{surface}/8\pi G \rfloor \cdot dA \tag{20}$$

That the question of collation of thermal radiation is best expressed as to how entropy can be collated into the nucleus of a big bang from innumerable places in space time, as the universe continually expands

Frankly, the author does not see how this is possible. Penrose (2007) in a 20 minute discussion with the author at the inaugural conference at the IGC Penn State university center, in 2007 stated that the necessary conditions for such a conformal mapping collation are contained within individual Black Hole Hawking temperature values. The author points to the dimensionality of the Black holes, and asks if a working conformal mapping is possible.

Penrose may well be right. The authors deliberations as to the inter relationship between entropy and Hawking temperature, as given by Eqn. (20) may reflect upon the author's limitations, not Penrose. If so, and if Penrose is right, the universe as we know it is far stranger than imagined by current cosmologists. In closing, the author wishes to point to how entropy is linked to IT models of information for the universe, as given by Lloyd (2002)

$$[\#operations] \leq S(entropy)/(k_B \cdot \ln 2) \sim 10^{120} \tag{21}$$

Parsing the connections between innumerable Black hole contributions to individual entropy into a grand value of entropy as given by Eqn. (21) below is currently beyond the Author's conceptual understanding at this juncture, but the concept is very intriguing and deserves serious study. In the conclusion below, the Author wishes to also high light what Dr. Smoot at the Challongue school talked about which is the growth of 'information' in a way parallel to Eqn. (21) above. Bosonification as mentioned below, where information is initially suppressed but released into cosmology evolution models at the onset of the Big bang favors the worm hole model as mentioned in this document.

# A modest proposal for future analysis: is the Worm hole from one prior universe to our present universe correct, or is Penrose right ?

As mentioned in the conclusion, Smoot talks about a maximum of up to $10^{120}$ bits of information in 'computing' processes in the present cosmological universe. There may be as little as up to only $10^8$ bytes of information going through the big bang itself. Two possibilities exist to explain this jump of information complexity

1. The worm hole approach suggested above leaves the direct possibility of encryption and , perhaps even using the initial worm hole itself as a way to 'zip file' most of the available information, in a suppressed, but orderly form, to be released after a Planck interval of time to blossom out to be re constituted.
2. What is suggested by Penrose would be a way of implying that one is taking an innumerable number of black holes, times up to $10^8$ bits of information per black hole , as a way to have innumerable parents of Hawkings radiation generated information which are re constituted in a mapping to a new singularity, to express a net conservation of information , in most ways, from a prior universe to our present, only in a different mix.

Penrose's proposal is intriguing because it is equivalent to the problem of biological information genetic transferal between one parent to the off spring. I.e. a mother may think that her son or daughter has not the same genetic transfer from one generation to the next, but the two parents together have a net genetic conservation law of enclosed 'bytes' of information maintained. In the Penrose picture, there may be several hundred thousand black holes of all shapes and sizes which contribute entropy largely due to Eqn (20) above, and then the total entropy from all those sources is maintained to produce , re constituted a net $10^{120}$ bites of information. The author would find it imperative to find experimentally falsifiable criteria to distinguish between these two paradigms of information storage, generation and retrieval.

# Conclusion

Eqn. (16) above outlines the current state of the art as to entanglement and its connections to the existence of a worm hole solution for transfer of vacuum energy from a prior universe to our own. In this document, we have alluded to the inescapable conclusion that entanglement in early universe geometry is intrinsically linked to the existence of short term quintessence in scalar fields, as seen in Beckwith (2007), and Moffat (2002)

This is important, for many reasons. One of them being that it explains the reason for why $w = -1$ so early in cosmological evolution. In addition the existence of $(\partial_0 \phi) \neq 0 \xrightarrow[t<Planck\ time]{} 0^+$ limit initially not zero is in tandem with the existence of casual discontinuity written up in another document where the existence of the time dependent Wheeler De Witt worm hole actually leads to causal discontinuity, which if quantum entanglement exists when $(\partial_0 \phi) \neq 0^+$ is in sync with the existence of faster than light transferal of thermal heat from a prior universe to our own

In a colloquium presentation done by Dr. Smoot (2007); he alluded to the following information theory constructions which bear consideration as to how much is transferred between a prior to the present universe in terms of information 'bits'. The following are Dr. Smoot's preliminary analysis of information content in the observable universe

1) Physically observable bits of information possibly generated in the Universe      - $10^{180}$
2) Holographic principle allowed states in the evolution / development of the Universe - $10^{120}$
3) Initially available states given to us to work with at the onset of the inflationary era  - $10^{10}$

4) Observable bits of information present due to quantum / statistical fluctuations  $-10^8$

Our guess is as follows. That the thermal flux so implied by the existence of a worm hole accounts for perhaps $10^{10}$ bits of information. These could be transferred via a worm hole solution from a prior universe to our present, as alluded to by Eqn. (9) above , and that there could be , perhaps $10^{120}$ minus $10^{10}$ bytes of information temporarily suppressed during the initial bozonification phase of matter right at the onset of the big bang itself . Then after the degrees of freedom dramatically drops during the beginning of the descent of temperature from about $T \approx 10^{32} \, Kelvin$ to at least three orders of magnitude less, as we move out from an initial red shift $z \approx 10^{25}$ to $T \approx \sqrt{\varepsilon_V} \times 10^{28} \, Kelvin \sim T_{Hawkings} \cong \frac{\hbar \cdot H_{initial}}{2\pi \cdot k_B}$ , as outlined by N. Sanchez .

The author wishes to state that he is convinced that the initial starting point of the Penrose model as far as inflation is correct.

What should we look forward to in the future? We should delineate more detail as to what would be transferred, possibly by entanglement information transfer from a prior universe , to our own, as well as understand how additional bytes of information came to be in the present Universe. All this would tie in with an accurate physical understanding of the points raised in the above section. And, in addition, the author is not ruling out in any fashion Dr. Penrose's (2007) very unorthodox suggestion as to the 'ripples' in the pond effect. The picture presented of worm hole transfer favors, implicitly, the big crunch models of cyclic universes. However, the author would be pleasantly surprised and intrigued if entropy collection of innumerable black holes, as related to by a grand value of entropy, as given by Eqn. (21), as mentioned by Lloyd (2002) and that Penrose (2007)is , indeed, verified as to his cyclic universe model.

It is worth mentioning that the author has done work in implementing Dowker's view of a causal discontinuity between prior and present universes, Beckwith, (2007) , implementing work done by Dowker ( 2003, 2007). It remains to be seen if such work is necessary in any way to compliment or round out Penrose's (2007) vision of a cyclic universe, minus the big crunch.

In any case, as mentioned above, the main task for future research would be in making a distinguishing set of experimental criteria which would allow astrophysical researchers to distinguish between an encryption of prior universe information to explain a drop between $10^{120}$ bites of information being reduced to $10^8$ bites of information and being re constituted via a worm hole 'zip file' encryption of information which would be subsequently released, and what appears to be the Penrose picture itself where there would be innumerable black holes , each contributing up to $10^8$ bites of information each, via Hawking radiation to conflate to after the big bang itself $10^{120}$ bites of information

# Bibliography

1) Barvinsky,A., Kamenschick,A., Yu, A. , " Thermodynamics from Nothing: Limiting the Cosmological Constant Landscape, Phy Rev D **74** 121502 (Rapid communications)
2) Beckwith, A., "Symmetries in evolving space time from present to prior universes", arXiv:math-ph/0501028. (2007)
3) Birrell, N., Davies, P.," Quantum fields in curved space", Cambridge Monographs on Mathematical physics , Cambridge , UK, 1984
4) Crowell, L.," Quantum Fluctuations of Space Time", World Scientific Series in Contemporary Chemical Physics, Volume 25, Singapore, Republic of Singapore, 2005
5) De Witt, B. " Quantum Gravity, the New Synthesis", in pp 680-745, of the volume " General Relativity , an Einstein Sentenary Survey", edited by Hawking, S. and Israel, W, Cambridge University Press, Cambridge, UK, 1979

6) Dowker, H. F., "Causal sets and the deep structure of spacetime", arXIV gr-qc/0508109v1 26 Aug 2005

7) Dowker, H.F., and Laflamme, R. in the paper "Worm holes and Gravitons" in Nuclear Physics B366 (1991) 209-232

8) Ida, D., Oda, K., and Park, S., " Rotating Black Holes at future Colliders : Greybody factors for brane fields", Phys rev D. **67**, 064025 (2003)

9) Kiefer, C.," Quantum aspects of Black Holes ", arXIV astro-ph/0202032 v1, February 1$^{st}$, 2002

10) Kolb, E., Turner, M.," The Early Universe:, Westview Press, 1994

11) Kurioukidis, A., Papadopoulos, D, " Pre – Inflation in the presence of Conformal Coupling", arXIV gr-qc/0401051 v1, January 13, 2004

12) Lloyd, S., "Computational capacity of the universe", Phys. Rev. Lett. 88, 237901 (2002)

13) Moffat, J. " Relativistic Causal Description of Quantum Entanglement", arXIV quant-ph/0204151 V 3 May 21$^{st}$, 2002

14) Park, D.K., Kim, H., and Tamarayan, S., "Nonvanishing Cosmological Constant of Flat Universe in Brane world Senarios," *Phys.Lett*. **B**535 (2002) pp. 5-10

15) Lecture notes taken at IGC conference, http://www.gravity.psu.edu/igc/conf_files/program_complete.pdf with respect to the Penrose presentation, " Conformal Cyclic cosmology, Dark Matter, and Black Hole Evaporation", plus questions asked of the lecturer in the aftermath of that presentation.

16) Lecture notes taken at 11 Paris Cosmology Colloquium, August 18, 2007 with respect Sanchez, N., "Understanding Inflation and Dark Energy in the Standard Model of the Universe"

17) Lecture notes taken at 11 Paris Cosmology Colloquium, August 18, 2007 with respect to Smoot, G, " CMB Observations and the Standard Model of the Universe"

18) Visser, M. " Lorentzian Wormholes, from Einstein to Hawking", AIP press, Woodbury, New York, USA, 1995